\documentclass{article}
\usepackage{spconf,amsmath,graphicx}

\usepackage{multirow,makecell,hyperref,pbox,booktabs}
\usepackage{float}
\usepackage{amsfonts}
\usepackage{comment}
\usepackage{graphics}
\usepackage{caption}
\usepackage{eso-pic}

\usepackage{fancyhdr}
\fancypagestyle{pageStyleOne}{%
    
    \fancyhf{}
    \fancyfoot[L]{\scriptsize © 20XX IEEE. Personal use of this material is permitted. Permission from IEEE must be obtained for all other uses, in any current or future media, including reprinting/republishing this material for advertising or promotional purposes, creating new collective works, for resale or redistribution to servers or lists, or reuse of any copyrighted component of this work in other works.}
}


\newcommand{\thickhline}{%
    \noalign {\ifnum 0=`}\fi \hrule height 1pt
    \futurelet \reserved@a \@xhline
}

\title{SpeakerNet: 1D Depth-wise Separable Convolutional Network for Text-Independent Speaker Recognition and Verification}
\name{Nithin Rao Koluguri, Jason Li, Vitaly Lavrukhin, Boris Ginsburg}
\address{NVIDIA, Santa Clara, USA }

\begin{document}

\maketitle
\thispagestyle{pageStyleOne}
\begin{abstract}
We propose SpeakerNet - a new neural architecture for speaker recognition and speaker verification tasks. It is composed of residual blocks with 1D depth-wise separable convolutions, batch-normalization, and ReLU layers. This architecture uses x-vector based statistics pooling layer to map variable-length utterances to a fixed-length embedding (q-vector). SpeakerNet-M is a simple lightweight model with just 5M parameters. It doesn't use voice activity detection (VAD) and achieves close to state-of-the-art performance scoring an Equal Error Rate (EER) of 2.10\% on the VoxCeleb1 cleaned and 2.29\% on the VoxCeleb1 trial files. 
\end{abstract}

\section{Introduction}
Speaker Recognition (SR) is a broad research area that solves two major tasks: speaker identification (who is speaking) and speaker verification (is the speaker whom they claim to be) \cite{Hansen2015}. In this work, we focus on the close, text-independent speaker recognition when the identity of the speaker is based on how speech is spoken, not necessarily in what is being said. Typically such SR systems operate on unconstrained speech utterances, which are converted to a vector of fixed length, called speaker embeddings. Speaker embeddings are also used in automatic speech recognition (ASR) \cite{zhao2018} and speech synthesis \cite{gibiansky2017}. 

In this paper, we propose SpeakerNet -- a new neural network-based architecture for the speaker recognition domain. SpeakerNet consists of three major parts: Encoder, Pooling layer, and Decoder. The Encoder is based on the QuartzNet architecture developed for ASR \cite{kriman2020QuartzNet}. It is composed of residual blocks, where each block consists of 1D depth-wise separable convolutions, batch norm, ReLU, and dropout layers. The Encoder converts audio of variable length into a sequence of acoustic features, which can be used to extract top level features, $E$. The Pooling layer maps the temporal sequence of acoustic features into a vector of fixed length by computing statistics on acoustic features. The Decoder, consisting of a series of fully connected layers, maps fixed-length vectors of dimension $D$ to a number of speakers $N$ to compute the probability that the current segment belongs to a speaker from the training set. This way the network extracts fixed-length representation from variable length speech segments. The network was trained end-to-end for speaker classification using the VoxCeleb1 and VoxCeleb2 dev \cite{Chung2018} datasets. 


The advantage of the proposed speaker embedding model is that it can be easily integrated with end-to-end deep ASR models \cite{kriman2020QuartzNet} since both models use the same architecture. \newline

The following are the main contributions of this paper:
\begin{enumerate}
    \item A new SR model: SpeakerNet based on the QuartzNet architecture with x-vector based pooling and without VAD
    \item Investigation of trade-off between training time, length of utterance, and Equal Error Rate
    \item Lighter model than current state-of-the-art (SOTA) model with similar performance
\end{enumerate}

The paper is organized as follows: In Section ~\ref{sec:related_work}, we review related work on Neural Network based speaker recognition. In Section \ref{sec:model_architecture}, we describe the QuartzNet architecture for speaker identification and the extraction of QuartzNet vectors (q-vectors) for speaker embedding. In Section \ref{sec:experiments}, we describe the training and evaluation methodology along with results on VoxCeleb1 trial files.

Pre-trained models and code are open sourced in NeMo, a conversational AI toolkit. \footnote{\href{https://github.com/NVIDIA/NeMo}{https://github.com/NVIDIA/NeMo}}

\section{Related work}
\label{sec:related_work}
Most traditional speaker recognition models use i-vectors \cite{dehak2010front}, based on Mel-Frequency Cepstral Coefficients (MFCC) which are used to build universal background models (UBM) \cite{reynolds2000}. A traditional UBM uses a Gaussian Mixture Model (GMM). One needs to learn a projection from the high-dimensional UBM space to the low-dimensional speaker i-vector. For speaker verification, \cite{Hansen2015} demonstrates how to use Probabilistic Linear Discriminant Analysis (PLDA) on i-vectors.

An alternative to GMM-UBM based speaker recognition is to use a Deep Neural Network (DNN)-based embedding. For example, Variani et al. \cite{Variani2014} trained a DNN to classify speakers at the development stage. The input of the DNN is formed by stacking the 40-dimensional log filter-bank features, together with wide temporal context -- 30 frames to the left and 10 frames to the right. The network was composed of four fully-connected max out \cite{Goodfellow13} layers, ReLU, pooling, and dropout layers \cite{Hinton2012}. During speaker enrollment, the trained DNN is used to extract speaker-specific features from the last hidden layer. The average of these speaker features termed d-vector, was used as the speaker model. At the evaluation stage, a d-vector is extracted for each utterance and compared to the enrolled speaker model to make a verification decision. The d-vector speaker verification system was still slightly worse than i-vector-based system, but it was more robust to noise. Heigold et al. \cite{heigold2016} proposed a d-vector-based model for \textit{text-dependent} speaker verification. The utterance-level speaker model was directly trained to compare pairs of embeddings. Switching from frame-level to utterance-level model and replacing fully-connected layers with a single LSTM layer significantly improved the verification accuracy.

Snyder et al. \cite{snyder2017,snyder2018} proposed x-vectors, a NN-based embedding  for \textit{text-independent} speaker verification. The model consists of 3 major blocks. The first block is the Time Delay Neural Network (TDNN) which works as a frame-level feature extractor. The input to this module is a variable-length sequence of acoustic features, e.g. MFCC. The x-TDNN is composed of 5 dilated convolutional layers. The second block is the `statistics pooling' layer, which receives the output of the final frame-level layer as input. This layer computes the mean and standard deviation for each channel over the input segment.  These segment-level statistics are concatenated together and passed to the last block, which has two hidden layers with dimension 512 and 300, either of which may be used to compute x-vectors. Finally, the network has a soft-max output layer. The NN is trained with cross-entropy loss to classify speakers. A probabilistic linear discriminant analysis (PLDA) classifier is used to compare x-vectors for same-or-different speaker decisions. X-vectors trained with significant data-augmentation outperformed i-vectors on multiple benchmarks and became a new baseline for speaker embeddings \cite{snyder2018}. 

Several extensions to x-vectors have been proposed recently. For example, a system from BUT \cite{zeinali2019description} achieved the best EER on VoxCeleb trial files with x-vector topology. They used a ResNet34 architecture consisting of 7M parameters to train for a speaker recognition task on different feature sets. Snyder et al. \cite{snyder2018} proposed end-to-end training with cross-entropy loss followed by fine-tuning for fewer epochs using additive angular margin loss. Another extension is based on the observation that some frames are more unique and important for discriminating speakers than others in a given utterance. One can add soft voice activity detection (VAD) or the  self-attention mechanism to calculate the weights for frame-level feature vectors, see \cite{Chowdhury2018},  \cite{wang2018}, \cite{Okabe2018}, \cite{Zhu2018}, \cite{shi2019}. 

There are several works showing performance of speaker embeddings with various losses like triplet loss \cite{bredin2017tristounet}, contrastive loss \cite{salehghaffari2018speaker}, and generalized end-to-end loss \cite{wan2018generalized}. Compared to other losses, systems trained with additive margin angular loss have shown better Equal Error Rate numbers \cite{coria2020comparison}.



\begin{figure}[!ht]
  \centering
  \centerline{\includegraphics[width=8.5cm]{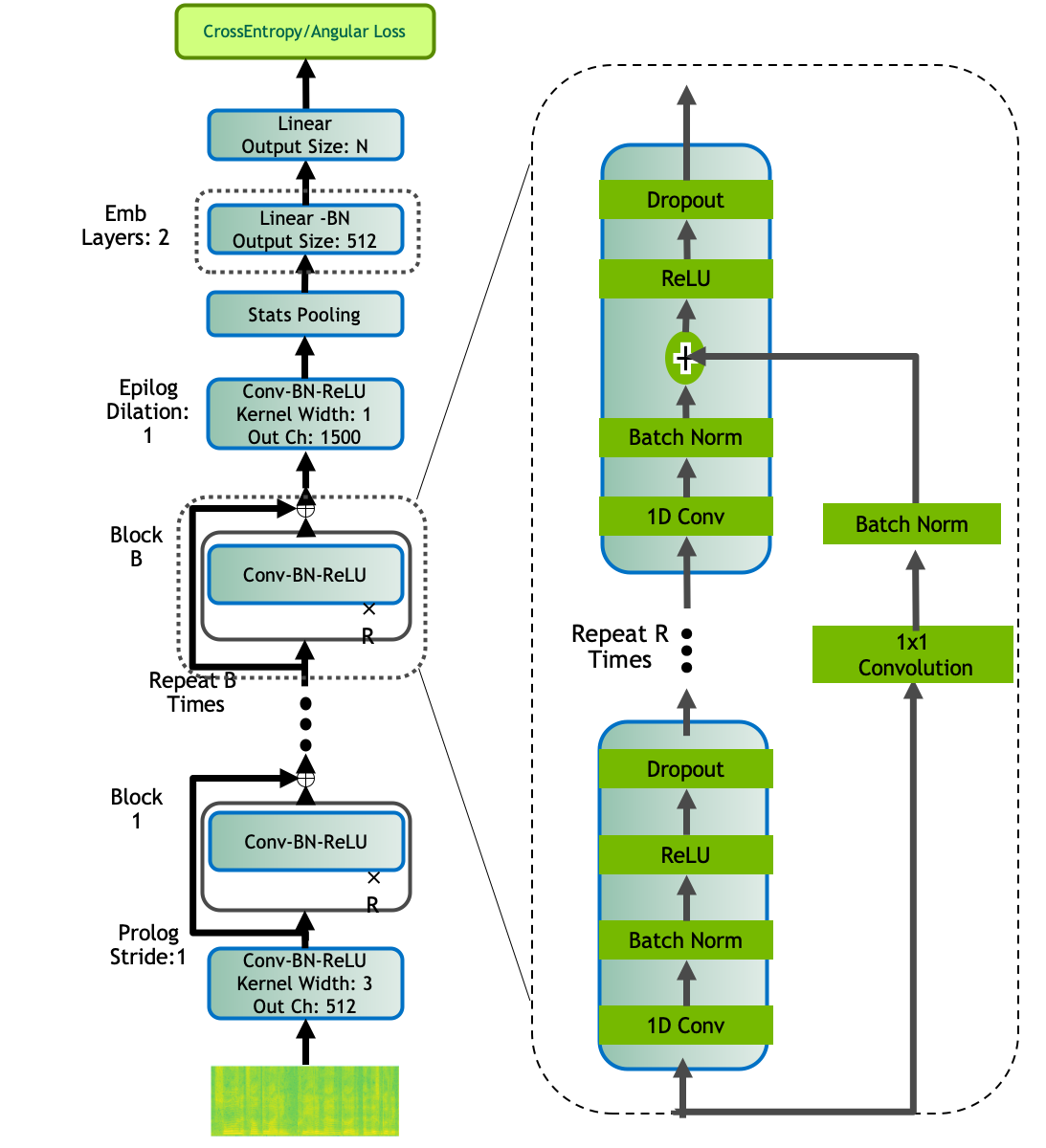}}
  \caption{SpeakerNet model architecture:  QuartzNet-based Encoder of size ${3 \times 2}$. Decoder Embedding combined from  two 512-vectors.}
  \label{fig:jasper_arch}
\end{figure}

\section{Model Architecture}
\label{sec:model_architecture}
\subsection{Encoder}
The model is based on the QuartzNet ASR architecture \cite{kriman2020QuartzNet} comprising of an encoder and decoder structure. We use the encoder of the QuartzNet model as a top-level feature extractor, and feed the output to the statistics pooling layer, where we compute the mean and variance across channel dimensions to capture the time-independent utterance-level speaker features.


The QuartzNet encoder used for speaker embeddings shown in Fig. \ref{fig:jasper_arch} has the following structure: a QuartzNet BxR model has B blocks, each with R sub-blocks. Each sub-block applies the following operations: a 1D convolution, batch norm, ReLU, and dropout. All sub-blocks in a block have the same number of output channels. These blocks are connected with residual connections. We use QuartzNet with 3 blocks, 2 sub-blocks, and 512 channels, as the Encoder for Speaker Embeddings (see 
Table~\ref{tab:JasperParams}). Note that all convolutional layers have stride 1 and dilation 1.

{\renewcommand{\arraystretch}{1.1}
\begin{table}
\caption{SpeakerNet-L Encoder 3x2x512: 3 blocks, each consisting of 2 1D-convolutional sub-blocks, plus 2 additional blocks.
}
\label{tab:JasperParams}
\centering
\scalebox{0.9}{
\begin{tabular}{c c c c c c} 
 \toprule
  \textbf {\# Blocks} & \textbf{Block} & \textbf{Kernel} & \textbf{\thead{\# Output\\Channels}} & \textbf{Dropout} & \textbf{\thead{\# Sub\\Blocks}} \\
 \midrule
 1 & Conv1 & 3 & 512 & 0.5 & 1 \\
 1 & B1 & 7 & 512 & 0.5 & 2 \\
 1 & B2 & 11 & 512 & 0.5 & 2 \\
 1 & B3 & 15 & 512 & 0.5 & 2 \\
 1 & Conv2 & 1 & 1500  & 0.0 & 1 \\
 \bottomrule
\end{tabular}
}
\end{table}}

\subsection{Decoder and Embeddings}
Top level acoustic Features, $E$ obtained from the output of encoder are used to compute intermediate features that are then passed to the decoder for getting utterance level speaker embeddings. The intermediate time-independent features are computed using a statistics pooling layer \cite{snyder2018}, where we compute the mean and standard deviation of features $E$ across time-channels, to get a time-independent feature representation $S$ of size $B \times 3000$. 

The intermediate features, $S$ are passed through the Decoder consisting of two layers each of output size 512 for a linear transformation from $S$ to the final number of classes $N$ for the larger ($L$) model, and a single linear layer of output size 256 to the final number of classes $N$ for the medium ($M$) model. We extract q-vectors after the final linear layer of fixed size $512, 256$ for SpeakerNet-L and SpeakerNet-M models respectively. 
\subsection{Loss function}
This SpeakerNet model was  trained end-to-end with both Cross-Entropy (CE) loss and additive angular margin (AAM) loss \cite{deng2019arcface}.

$$L_{AAM} = -\frac{1}{N} \sum_{i=1}^{N} \log \frac{e^{s\left(\cos \left(\theta_{y_{i}}+m\right)\right)}}{e^{s\left(\cos \left(\theta_{y_{i}}+m\right)\right)}+\sum_{j=1, j \neq y_{i}}^{n} e^{s \cos \theta_{j}}}$$
where $m$ is margin, $s$ is scale and $\theta_{j}$ is the angle between the final linear layer weight $W_{j}$ and incoming feature $x_{i}$. Here $m$ and $s$ are predefined hyper parameters. 

\section{Experiments}
\label{sec:experiments}
\subsection{Dataset}
We use the VoxCeleb1 and VoxCeleb2 dev dataset to train SpeakerNet. To augment the training data we use the MUSAN and RIR impulse corpora. In our experiments, we ignore the speech subpart of the MUSAN corpus. Dataset statistics of VoxCeleb1 and VoxCeleb2 are provided in Table \ref{tab:datasets}.


\begin{table}
\centering
\caption{VoxCeleb1 and VoxCeleb2 dev dataset statistics}
\label{tab:datasets}
\resizebox{0.48\textwidth}{!}{%
\begin{tabular}{cccccc}
\hline
\multirow{2}{*}{\textbf{Dataset}} & \multirow{2}{*}{\textbf{\# Speakers}} & \multicolumn{2}{c}{\textbf{Duration} \textit{(s)}} & \multicolumn{2}{c}{\textbf{\# Utterances}} \\
  & & min & max & min & max \\ 
 \hline
VoxCeleb1 dev & 1211 & 232 & 7870 & 41 & 902 \\
VoxCeleb2 dev & 5994 & 88 & 8678 & 19 & 450 \\ 
 \hline
\end{tabular}
}
\end{table}


\begin{figure}
    \centering
    \includegraphics[width=8cm]{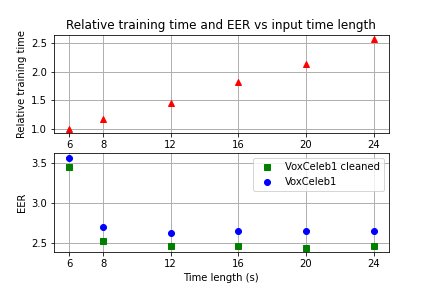}
    \caption{EER  and relative training time for MUSAN-augmented systems depending on input time length}
    \label{fig:training_time}
\end{figure}

\begin{table*}[!ht]
\centering
\caption{EER for SpeakerNet-L with various data augmentations}
\label{tab:augmentaion}
\begin{tabular}{c c c c}
\hline
 Augmentation  & VoxCeleb1 cleaned, EER (\%) & VoxCeleb1, EER (\%) \\
\hline
   $-$            &  $2.85$     & $2.99$    \\ 
          MUSAN            &  $\textbf{2.53}$    & $\textbf{2.7}$    \\
           RIR           &  $2.64$   & $2.8$    \\ 
         MUSAN + RIR          &   $2.89$   & $3.02$    \\ 
\hline
\end{tabular}
\end{table*}

\begin{table*}[!ht]
\centering
\caption{Comparison  of SpeakerNet-M and  SpeakerNet-L  with SOTA speaker recognition  systems}
\label{tab:results}
\begin{tabular}{ c c c c c c c}
\hline
System & \#Parameters (M) & Embedding Size & Loss & VoxCeleb1 cleaned, EER (\%) & VoxCeleb1, EER (\%) \\
\hline
Kaldi \cite{snyder2018}  &  9      & 512            & CE    &  -     & $3.10$    \\ 
BUT \cite{zeinali2019description}  &  7     & 256            & AAM    &  $\textbf{1.22}$     & -    \\
\hline
SpeakerNet-L  &  8     & 512        & AAM    &  $2.10$     & $2.32$    \\ 
SpeakerNet-M  &  \textbf{5}     & 256        & AAM    &  $2.14$     & $\textbf{2.29}$ \\

\hline
\end{tabular}
\end{table*}

\subsection{Experiment Setup}
Every speaker recognition experiment consists of common data pre-processing steps for training, development, and evaluation steps. During pre-processing we don't use VAD to avoid dependence on additional model in order to simplify the pipeline. We compute acoustic features from speech recordings for every 20 ms frame window shifted over 10 ms. The acoustic features are 64-dimensional Mel-Frequency Cepstral Coefficients (MFCCs), computed from spectrograms calculated using a 512 FFT size and a Hann window. Normalization over the frequency axis is then performed over the MFCCs. Every utterance fed to the encoder are of sizes $T \times 64$, where $T$ is the number of frames in a given speech utterance file. 

Unless specified otherwise, all systems in Tables \ref{fig:training_time} and \ref{tab:augmentaion} are SpeakerNet-L models, which have been trained for 200 epochs with SGD optimizer, with initial learning rate $0.006$ using cosine annealing learning rate schedule. All the EER evaluations were done using a cosine similarity back-end.

\subsection{Results}
\label{sec:results}
The performance of the SpeakerNet system on VoxCeleb1 and VoxCeleb1 cleaned trial with various experiments are shown in Tables \ref{tab:augmentaion}, \ref{tab:results}. In all these experiments we train our model on the combined VoxCeleb1 and VoxCeleb2 dev datasets. 

We train these systems initially as a speaker recognition model with 10 percent of audio files of each speaker set aside as validation data from training VoxCeleb1 and VoxCeleb2 dev sets. All the systems mentioned in Fig. \ref{fig:training_time} and Tables \ref{tab:augmentaion}, \ref{tab:results} are trained with the same QuartzNet $3\times2\times512$ encoder architecture and with a decoder of intermediate linear with shapes 512 and 512, except SpeakerNet-M model where it has only one intermediate linear layer of size 256.

In Table \ref{tab:augmentaion}, we compare the performance of SpeakerNet-L with and without augmented data that are trained end-to-end using cross-entropy loss. We observe that augmentation with additive MUSAN noise improves EER significantly on VoxCeleb sets. As expected RIR augmentation does not help much for these test sets which consist of near field recordings but should improve performance on far-field data.

With this dataset setup, we trained SpeakerNet-L end-to-end using additive margin angular loss. We observed a high degree of sensitivity on validation curves with slight variations in the margin $m$ and scale $s$ for angular loss. Out of various range of values we experimented, $m=0.2$ and $s=30$ performed better with EER of 2.10\%. Systems mentioned in Table \ref{tab:augmentaion} and \ref{tab:results} were all trained with audio samples of less than or equal to 8 seconds, if input audio is more than 8 seconds we randomly chunk the input audio to a 8 second segment. This time length was chosen in order to cut down training time by 2x with negligible performance degradation when compared to a system with all the training data sent as is. In Fig. \ref{fig:training_time}, we show the performance of systems with this variable "time length (s)" ranging from 6 to 20 seconds in steps of 4. As can be observed, EER decreases at a very small rate which is not comparable to increase in training time. 

Finally we compare our two best models against SOTA systems in Table \ref{tab:results}. SpeakerNet-M is the most lightweight model having only 5M parameters, 256 embedding size that doesn't use VAD. At the same time, it performs better on VoxCeleb1 trials with smaller EER numbers than larger Kaldi system but worse than SOTA BUT system.


\section{Conclusions}
We have presented a new neural architecture, SpeakerNet, for end-to-end speech verification and identification tasks. The proposed SpeakerNet-M has fewer parameters when compared to SOTA, doesn't use VAD as a pre-processing step, and was trained on smaller sample lengths in order to reduce training time. In spite of all the cut offs we made, our model shows very similar performance on VoxCeleb1 trial files when compared to SOTA systems. The models' implementation is available in \url{https://github.com/NVIDIA/NeMo} repository under Apache 2.0 license. And we also provide pre-trained checkpoints.

In our future work, we plan to use these systems to develop an end-to-end speaker diarization system for better diarization error rates.  

\section{Acknowledgments}
We would like to thank NVIDIA AI Applications team for the help and valuable feedback.

\bibliographystyle{IEEEbib}
\bibliography{speakerbib}

\end{document}